# A Hierarchical Semantic Overlay for P2P Search

Tao Gu [1,2], Hung Keng Pung [2], Daqing Zhang [1]

[1] Institute for Infocomm Research, 21 Heng Mui Keng Terrace, Singapore

[2] National University of Singapore, 3 Science Drive 2, Singapore

tgu@i2r.a-star.edu.sg; punghk@comp.nus.edu.sg; daqing@i2r.a-star.edu.sg

*Abstract* — **In this paper, we propose a hierarchical semantic overlay network for searching heterogeneous data over wide-area networks. In this system, data are represented as RDF triples based on ontologies. Peers that have the same semantics are organized into a semantic cluster, and the semantic clusters are self-organized into a one-dimensional ring space to form the top-level semantic overlay network. Each semantic cluster has its low-level overlay network which can be built using an unstructured overlay or a DHT-based overlay. A search is first forwarded to the appropriate semantic cluster, and then routed to the specific peers that hold the relevant data using a parallel flooding algorithm or a DHT-based routing algorithm. By combining the advantages of both unstructured and structured overlay networks, we are able to achieve a better tradeoff in terms of search efficiency, search cost and overlay maintenance cost.**

## I. INTRODUCTION

Peer-to-Peer (P2P) systems have recently become more and more popular to share and exchange data over wide-area networks. Unstructured P2P systems such as Gnutella [1] and Kazza [2] allow peers to interconnect freely, making it easy to handle the dynamic changes of peers and their data. These systems do not impose any structure on the data, and hence have low overlay maintenance cost. However, a search has to be flooded to all peers in the network including those peers that do not have relevant data. Although flooding is simple and works well in a dynamic environment, it will generate large amount of redundant messages, which makes the system not scalable. Structured P2P systems such as Chord [3], CAN [4], and Pastry [5] typically use distributed hash tables (DHTs) to direct a search request to the specific peers. They can guarantee completing search in a logarithmic number of steps. However, data placement and network topology in these systems are tightly controlled based on distributed hash functions, which results in high overlay maintenance cost for updating the relevant information in the overlay network.

Recently, hybrid P2P systems such as pSearch [6] and SSW [7] combine the advantages of both unstructured and structured P2P systems, and create a better possibility of building large-scale yet reliable P2P systems. Inspired by these systems, we propose a hierarchical semantic overlay network for searching heterogeneous data over wide-area networks. The basic idea is to construct a two-level semantic P2P overlay network based on metadata (i.e., ontology), which is essentially a hybrid approach, to facilitate efficient search. We use RDF [8] based data model to describe resources because RDF data are machine-understandable and machine-processable, and RDF has been widely accepted as a standard way for representing and exchanging information on the web.

In the top-level semantic overlay, peers that have the same semantics are grouped into a cluster called semantic cluster, according to ontologies. These semantic clusters are self-organized into a certain structure, i.e., in our case a one-dimensional ring space. Different overlaying techniques could be used for connection of peers in each semantic cluster. Since the peers within a semantic cluster usually have the same characteristic in term of type of data and their dynamicity, we can organize them using the same P2P overlay, i.e., either an unstructured overlay or a DHT-based overlay. For example, in the ubiquitous computing domain, for the peers of a semantic cluster that have more dynamic data such as users' location information, we can use a Gnutella-like overlay; and for the peers that have static data such as users' profile information, we can use a DHT-based overlay. By separating the semantics of peers and constructing an appropriate overlay for each semantic cluster, we are able to achieve a better tradeoff in terms of search efficiency, search cost and overlay maintenance cost.

The contributions of this paper are: (i) an ontology-based approach to hybrid P2P systems, where heterogeneous data are self-organized into a two-level semantic overlay network, for facilitating efficient P2P search; (ii) a novel ontology-based semantic clustering technique; and (iii) a DHT-based inter-cluster routing algorithm, and a parallel flooding algorithm for an unstructured overlay network.

## II. TWO-LEVEL SEMANTIC OVERLAY NETWORK

In our system, a large number of peers are self-organized into a two-level semantic overlay network, and grouped in accordance with their semantics. Each peer maintains a local data repository which includes RDF data instances and their corresponding ontologies, and supports an RDF-based query using RDQL [9]. Upon creation, each peer will first extract the semantics of its data, and then join an appropriate semantic cluster. These semantic clusters have logically interconnected as a one-dimensional ring structure based on the small world model [10] to form the top-level overlay. In low-level overlays, peers in each semantic cluster can further be arranged as an unstructured overlay or a DHT-based overlay. Upon receiving a query, the peer first processes it to obtain the semantic cluster that the query is associated with; it then routes the query to an appropriate semantic cluster. Within a semantic cluster, a peer may route that query using a parallel flooding algorithm or a DHT-based algorithm. Peers that receive the query do a local search, and return results. Figure 1 provides an overview of our proposed two-level semantic overlay network.

This joint work is supported by Institute for Infocomm Research and National University of Singapore. Copyright is held by the authors.

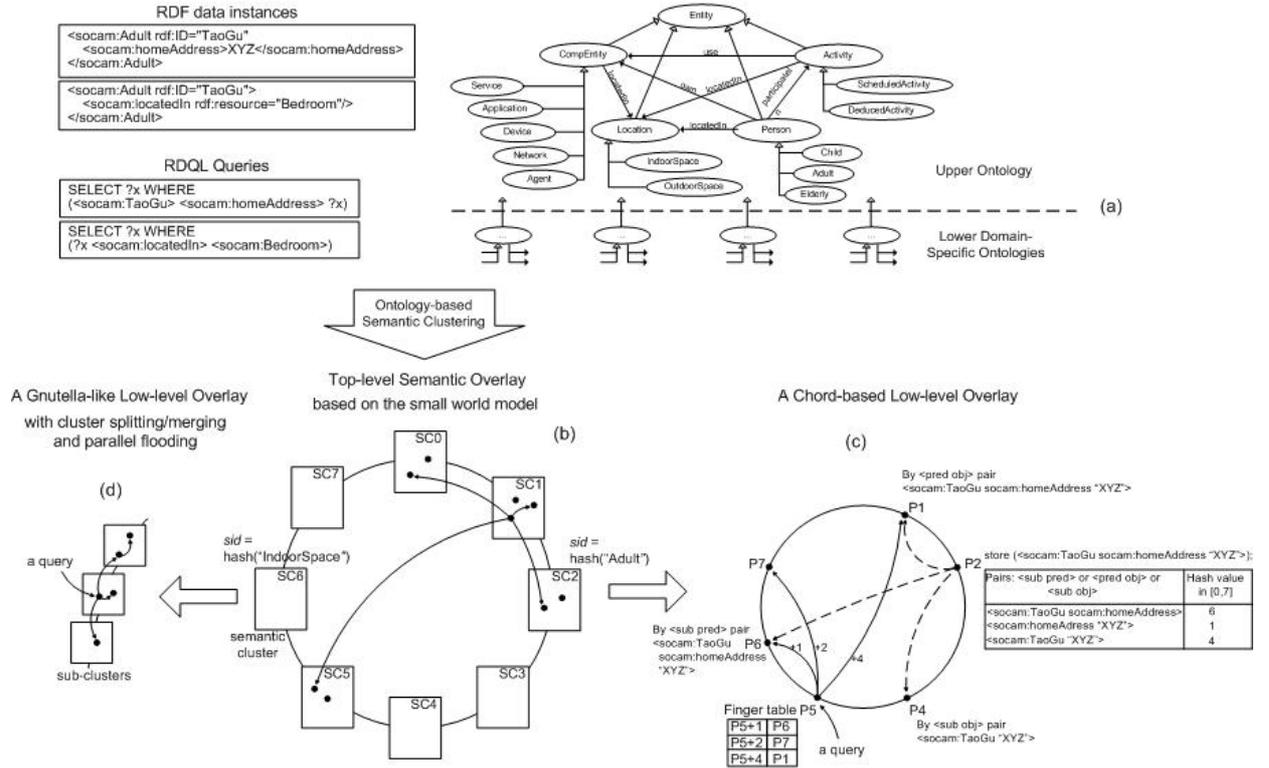

Figure 1. Overview of the two-level semantic overlay network

## A. Ontology-based Semantic Clustering

To extract the semantics of RDF data and RDQL queries, we propose an ontology-based semantic clustering technique. As compared to other semantic extraction techniques such as VSM [11] and LSI [12] used in pSearch, the formal design of ontologies minimizes the problems of synonyms and polysemy incurred by VSM. Based on ontologies, RDF data can be mapped to appropriate semantic cluster(s) accurately and quickly without costly computation as in LSI.

Our clustering technique makes use of ontology. We adopt a two-level hierarchy in the ontology design. The upper ontology defines common concepts in a computing domain or an application domain, and it is shared by all peers. Each peer can define its own concepts in its lower ontology; different peers may have different sets of lower ontologies. An example of such design in the ubiquitous computing domain is shown in Figure 1(a). The leaf classes in the upper ontology are used as semantic clusters, and each semantic cluster will be assigned with a unique ID upon its creation in the system.

The clustering computation is done locally at each peer. Let $SCn_{sub}$, $SCn_{pred}$, $SCn_{obj}$ where $n = 1,2,...$ denote the semantic clusters extracted from the subject, predicate and object of a data triple respectively. If the predicate of an RDF data triple is of type `ObjectProperty`, we obtain the semantic cluster using $(SC1_{pred} \cup SC2_{pred} \cup ... SCn_{pred}) \cap (SC1_{obj} \cup SC2_{obj} \cup ... SCn_{obj})$. If the predicate of an RDF data triple is of type `DatatypeProperty`, we obtain the semantic cluster using $(SC1_{sub} \cup SC2_{sub} \cup ... SCn_{sub}) \cap (SC1_{pred} \cup SC2_{pred} \cup ... SCn_{pred})$. For example, in Figure 1, RDF triple `<socam:TaoGu socam:locatedIn socam:Bedroom>` can be mapped to $IndoorSpace \cap IndoorSpace = IndoorSpace$. An RDQL query follows the similar procedure to obtain its semantic cluster.

## B. Top-level Overlay and Inter-cluster Routing

A peer will be assigned with an ID upon joining the system. We use SHA1 hash function to generate peers' identifier space. To incorporate semantic information associated with a peer, we dedicate part of hashed peer identifiers to correspond to its semantic cluster. More specifically, in a $k$-bits identifier space, we allocate $m$-bits for semantic cluster and $n$-bits for its IP address, where $k = m + n$. An example of a peer's ID generated by hashing its semantic cluster `Adult` and its IP address `137.132.74.135` is given below:

`peer id = [hash_m("Adult")][hash_n("137.132.74.135")]`

With this encoding scheme, we are able to construct a two-level overlay network and identify a peer uniquely in the system. The first $m$-bits of a peer's ID (which are called *Semantic Cluster ID* or *sid* in short) corresponds to a semantic cluster in the top-level overlay, and the last $n$-bits represents the peer's ID in the low-level overlay. As a peer may obtain multiple semantics from its local data, we choose the semantic cluster corresponding to the largest set of data to place the peer. For data corresponding to other semantic clusters, the peer registers the indices of these data to the respective semantic cluster.

We follow the small world model to construct the top-level overlay. Studies show that searches can be efficiently routed in small world networks when each peer knows its local neighbors (called short-range contacts), each peer also knows a small number of randomly chosen distant peers (called long-range contacts) with probability proportional to *1/d* where *d* is the distance [10]. The constant number of contacts and small average path length serve as the motivation for us to build the top-level overlay using the small world model. For example, as shown in Figure 1(b), a peer in *SC1* maintains a short-range contact to a peer in its own semantic cluster, a short-range contact to a peer in *SC0* and *SC2* respectively, and a long-range contact to a peer in *SC5*. Long-range contacts aim at providing shortcuts to reach other semantic clusters quickly; via short-range and long-range contacts, a search in the top-level overlay can be guided greedily by comparing *sids* of the destination and the traversed peers.

Query routing involves two steps: inter-cluster routing and intra-cluster routing. A query will be first forwarded to an appropriate semantic cluster in the top-level overlay and routed to destination peers in the low-level overlay. Upon receiving a query, a peer first extracts the associated semantic cluster, and obtains the search key by hashing the semantic cluster. It then compares the search key with the most significant *m*-bits of its neighbors' identifiers, and forwards the query to the closest neighboring peer. This forwarding process is recursively carried out until the destination semantic cluster is reached. We shall describe the intra-cluster routing in the next section.

*C. Low-level Overlays and Intra-cluster Routing*

In low-level overlays, peers in each semantic cluster can be organized as an unstructured overlay or a DHT-based overlay. In our system, we use a Gnutella-like overlay to arrange the peers in a semantic cluster whose data are more dynamic, for example, users' location information; and use a DHT-based overlay such as Chord or CAN to arrange the peers in a semantic cluster whose data are more static, for example, users' profile information.

For an illustration, we use Chord to construct the low-level overlay for semantic cluster *SC2*, and show a 3-bit Chord identifier space of 6 peers in Figure 1(c). We build distributed indices for each RDF data triple by storing each triple three times in Chord. This is achieved by applying the hash function to the `<sub pred>`, `<pred obj>` and `<sub obj>` pairs to generate the keys for each data triple. Each data triple will be stored at the successor peers of the hashed key values of these pairs. Figure 1(c) illustrates the process that peer *P2* stores the tripe `<socam:TaoGu socam:homeAddress "XYZ">` in Chord. For intra-cluster routing, when a query reaches a peer in a Chord-based semantic cluster, the peer will use its finger table to route the query. We show peer *P5*'s finger table in Figure 1(c). For example, when receiving an RDQL query `SELECT ?x WHERE (<socam:TaoGu> <socam:homeAddress> ?x)`, peer *P5* looks up the hashed `<sub pred>` pair using its fingers. The query will then be forwarded to peer *P6*, and the result `<socam:TaoGu socam:homeAddress "XYZ">` will be returned finally.

In Figure 1(d), we also show the use of a Gnutella-like overlay to construct the low-level overlay for semantic cluster *SC6*. To improve the blind flooding mechanism used in a Gnutella-like overlay, we propose a parallel flooding mechanism that is achieved by splitting a semantic cluster. If a semantic cluster exceeds the maximum cluster size – a pre-defined value, it will be split into two sub-clusters. During splitting, we use a proximity-based grouping technique in which close-by peers are grouped into the same sub-cluster to ensure the overlay topology matches the underlying physical topology. If the size of a sub-cluster falls to the minimum size due to peer leaving, it will be merged with one of its neighboring sub-clusters. For routing in a Gnutella-like semantic cluster, a peer first forwards the query to its neighboring sub-clusters, and then floods it to its own sub-cluster. This process is recursively carried out until all the sub-clusters corresponding to the query's semantic cluster are covered. In this way, search is done in parallel within a semantic cluster, and hence search efficiency can be improved as compared to the blind flooding.

III. EVALUATION AND ON-GOING WORK

We use simulation to evaluate our proposed two-level semantic overlay network. We have implemented the ontology-based semantic clustering, and built the top-level overlay and a Gnutella-like low-level overlay. The initial evaluation results look promising; they have shown that the proposed parallel search and clustering techniques significantly improve the search efficiency in a Gnutella-like overlay. The overlay maintenance cost is low, and the system adapts to the peer dynamics quickly. We omit the results due to space constraint.

We are working on a Chord-based low-level overlay and studying its performance. We will also evaluate the overall system performance in terms of search efficiency, search cost and overlay maintenance cost by having heterogeneous data with different level of dynamicity in the system. We also plan to build typical applications in various application domains to demonstrate the wider applicability of our system in a real world environment.

REFERENCES

[1] Gnutella, http://gnutella.wego.com
[2] Kazza, http://www.kazza.com
[3] I. Stoica, R. Morris, D. Karger, F. Kaashoek, and H. Balakrishnan. Chord: A Scalable Peer-to-Peer Lookup Service for Internet Applications. In Proc. of ACM SIGCOMM, 2001.
[4] S. Ratnasamy, P. Francis, M. Handley, R. Karp, and S. Shenker. A Scalable Content Addressable Network. In Proc. of ACM SIGCOMM, 2001.
[5] A. Rowstron and P. Druschel, Pastry: Scalable. Distributed Object Location and Routing for Large-scale Peer-to-Peer Systems. Lecture Notes in Computer Science, 2218:161–172, November 2001.
[6] C. Q. Tang, Z. C. Xu, and S. Dwarkadas. Peer-to-Peer Information Retrieval Using Self-Organizing Semantic Overlay Networks. In Proc. of ACM SIGCOMM 2003, Karlsruhe, Germany, August 2003.
[7] M. Li, W. C. Lee, and A. Sivasubramaniam. Semantic Small World: An Overlay Network for Peer-to-Peer Search. In Proceedings of the International Conference on Network Protocols, October, 2004.
[8] RDFStore. http://rdfstore.sourceforge.net.
[9] RDQL, http://www.w3.org/Submission/2004/SUBM-RDQL-20040109/
[10] J. Kleinberg. The Small-World Phenomenon: an Algorithm Perspective. In Proc. of the 32nd ACM Symposium on Theory of Computing, 2000.
[11] M. Berry, Z. Drmac, and E. Jessup. Matrices, Vector Spaces, and Information Retrieval. SIAM Review, 41(2):335–362, 1999.
[12] S. C. Deerwester, S. T. Dumais, T. K. Landauer, G. W. Furnas, and R. A. Harshman. Indexing by Latent Semantic Analysis. Journal of the American Society of Information Science, 41(6):391–407, 1990.